\documentclass{article}

     \usepackage[nonatbib, final]{tackling_climate_workshop_style}

\usepackage[utf8]{inputenc} 
\usepackage[T1]{fontenc}    
\usepackage{hyperref}       
\usepackage{url}            
\usepackage{booktabs}       
\usepackage{amsfonts}       
\usepackage{nicefrac}       
\usepackage{microtype}      
\usepackage{graphicx}

\usepackage[backend=biber,style=numeric,sorting=ynt]{biblatex}
\usepackage{xcolor}

\definecolor{vermilion}{rgb}{0.89, 0.26, 0.2}

\title{Multi-Resolution Analysis of the Convective Structure of Tropical Cyclones for Short-Term Intensity Guidance}

\author{%
  Elizabeth Cucuzzella \\
  Department of Statistics \\and Data Science\\
  Carnegie Mellon University\\
  \texttt{ecucuzze@andrew.cmu.edu} \\
  \And
  Tria McNeely \\
  Department of Statistics \\and Data Science \\
  Carnegie Mellon University \\
  \texttt{triamcneely@gmail.com} \\
  \AND
  Kimberly Wood \\
  Department of Atmospheric Sciences \\ and Hydrology \\
  University of Arizona\\
  \texttt{kimwood@arizona.edu} \\
  \And
  Ann B. Lee \\
  Department of Statistics \\and Data Science \\
  Carnegie Mellon University\\
  \texttt{annlee@andrew.cmu.edu} \\
}

\begin{document}

\maketitle

\begin{abstract}
 Accurate tropical cyclone (TC) short-term intensity forecasting with a 24-hour lead time is essential for disaster mitigation in the Atlantic TC basin. Since most TCs evolve far from land-based observing networks, satellite imagery is critical to monitoring these storms; however, these complex and high-resolution spatial structures can be challenging to qualitatively interpret in real time by forecasters. Here we propose a concise, interpretable, and descriptive approach to  quantify fine TC structures with a multi-resolution analysis (MRA) by the discrete wavelet transform, enabling data analysts to identify physically meaningful structural features that strongly correlate with rapid intensity change. Furthermore, deep-learning techniques can build on this MRA for short-term intensity guidance.
\end{abstract}

\section{Introduction}
Since 1980, tropical cyclones (TCs) have cost an average of \$23 billion in damages per weather event in the United States, resulting in more than \$1.5 trillion in damages~\cite{NOAAcosts}. As a result of rising temperatures, it has become increasingly urgent to forecast short-term intensity changes with a 24-hour lead time~\cite{AdvancedML}. In particular, rapid intensification (RI) events, defined as an increase of at least 30 knots in maximum sustained wind speeds within a 24-hour period~\cite{Definitions}, can pose a great threat to life and property as their rapid fluctuations make them difficult to prepare for. 

The convective structure of a TC has long been associated with its intensity; methods such as the Dvorak technique identify that greater temperature differences in the storm eye and cloud coverage are generally associated with strong TCs~\cite{DvorakTechnique}. Infrared (IR) images serve as a proxy for the convective structure of a cyclone because IR infers cloud-top temperature via black-body assumptions. Whereas modern machine-learning algorithms can improve TC forecasts based on massive complex data, end-to-end AI algorithms have a black-box feel, making it hard for forecasters to gain physical insight from intensity forecasts. Previous work has shown that sequences of radial profiles or principal components of IR images can be used to identify storms structurally primed for RI, as well as provide interpretability by highlighting structural archetypes preceding rapid intensity changes~\cite{StructuralForecasting, Improvement}. However, these summary statistics potentially lose critical directional and finer-scale information about the spatio-temporal evolution of the TC~\cite{GOESUnlocking}. 

We propose a richer, multi-resolution approach (MRA) to convective structure analysis which uses an orthogonal wavelet decomposition to obtain sparse representations of the initial sequence of high-resolution IR images. The challenge in using the full IR image for forecasting arises from the inherent noisiness of satellite imagery and the demand for greater computational complexity. Here we suggest a sparse, efficient representation of the initial IR image at multiple resolutions to capture both coarse and fine information about a TC's convective structure. We then apply machine learning algorithms in the wavelet space to classify sequences of images as structurally primed for RI. The sparse nature of the wavelet representation allows for interpretable analysis of spatio-temporal convective patterns that are correlated with rapid intensity changes. An additional benefit of the sparse MRA lies in the ability to propagate a reduced set of wavelet coefficients forward in time via generative models~\cite{FirstWaveletProp, WaveletProp}, allowing for interpretable structural forecasting and improved TC intensity guidance.

\section{Data}
The Geostationary Operation Environmental Satellites (GOES) provide high spatial ($\leq4$km) resolution imagery of the Atlantic TC basin~\cite{GOES}. In our preliminary study, we use 24-hour sequences of long-wave infrared (IR $\sim10.7 \mu$m) imagery. The intensity of the storms are drawn from the NHC's HURDAT2 database~\cite{HURDAT}. Data pertaining to environmental variables affecting TC intensity, such as atmospheric moisture and vertical wind shear, are sourced from inpute data for the Statistical Hurricane Intensity Prediction Scheme (SHIPS) which is an operational statistical-dynamical model used for intensity and track forecasting, with values provided every 6 hours~\cite{SHIPS}.

\section{Multi-Resolution Convective Structure Analysis for TC Intensity Guidance}
In this section, we show preliminary results for how a wavelet representation of TC satellite imagery can enhance interpretability as well as support improvements to intensity predictions as compared to the original satellite imagery.

\subsection{Discrete Wavelet Transform for Representing Convective Structure} 
Wavelets~\cite{WaveletBook} allow us to conveniently decompose images into structures on different scales. In our preliminary analysis, we express each GOES IR image $f(x)$ as a linear combination $f(x)=\Sigma_{j,n,k}c^k_{j,n}\psi^k_{j,n}(x)$ with $\psi^k_{j, n}(x)=2^{-j}\psi^k(2^{-j}x-n)$, $\forall j \in \mathbb{Z}^+, n \in \mathbb{Z}^2, 1\leq k \leq 3$~\cite{MallatBook}, where $\psi^k_{j,n}$ are orthogonal Debauchies wavelet functions at scale $j$ (for a square of width $2^j$) in the horizontal, vertical, or diagonal direction (indexed by $k$). With a fast discrete wavelet transform~\cite{PyWavelet}, we encode directional changes of convective structure at both fine and coarse resolutions. In our preliminary analysis, we chose to study structures at scales  $j=1,2,3$. Because we are only interested in the regions with largest gradients in convective structure, we further threshold the wavelet coefficients to only retain the top 10\% highest magnitude coefficients. In addition, because TC intensity is most strongly correlated with temperature gradients in and near the eye~\cite{DvorakTechnique}, we only retain coefficients within the inner 25\% of radii at the finest scale $j=1$. With these cuts, we achieve a total compression of the original image by a factor of $\frac{1}{20}$. 

\begin{figure}
  \centering
  \textnormal{ROC Curves}\par\medskip
  \includegraphics[width=0.45\linewidth]{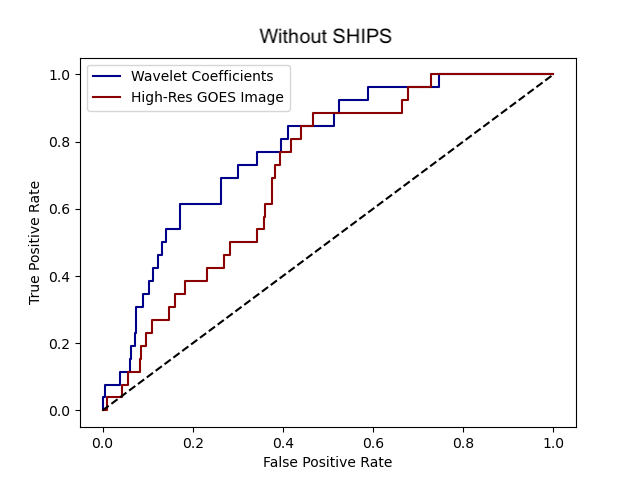}
  \includegraphics[width=0.45\linewidth]{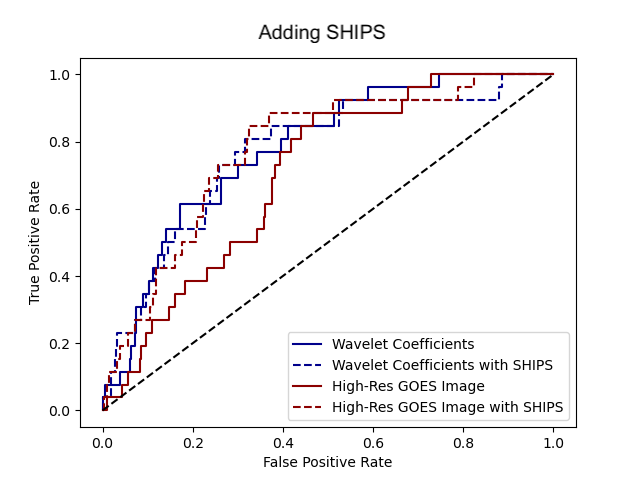}
  \caption{\small {\em{Left:}} CNNs trained with our wavelet coefficients (solid \textcolor{blue}{blue}) lead to much better TC intensity predictions, as indicated by a larger area under the ROC curve, than when working directly with the original GOES imagery (solid \textcolor{red}{red}). {\em{Right:}} Adding SHIPS variables marginally improves the wavelet results (dashed \textcolor{blue}{blue}), but is necessary for GOES (dashed \textcolor{red}{red} after adding SHIPS) to be competitive with wavelet results without SHIPS (solid \textcolor{blue}{blue}).}
  \label{fig:t0}
\end{figure}

\subsection{Sequence Classification for Nowcasting TC intensities} 
\begin{figure}
    \centering
    \textnormal{Hurricane Eta (2020)}\par\medskip
    \includegraphics[width=0.45\linewidth]{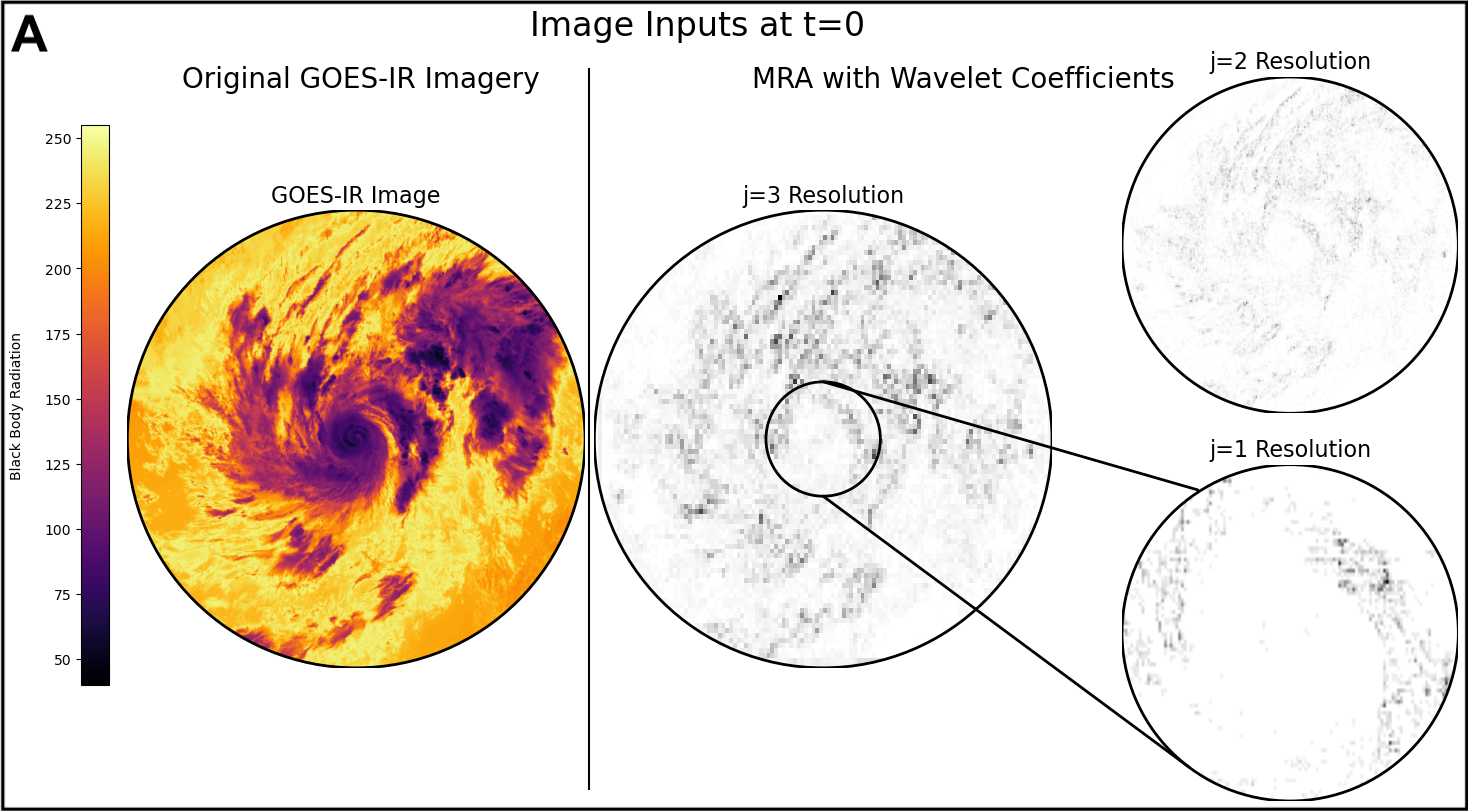}
    \includegraphics[width=0.46255\linewidth]{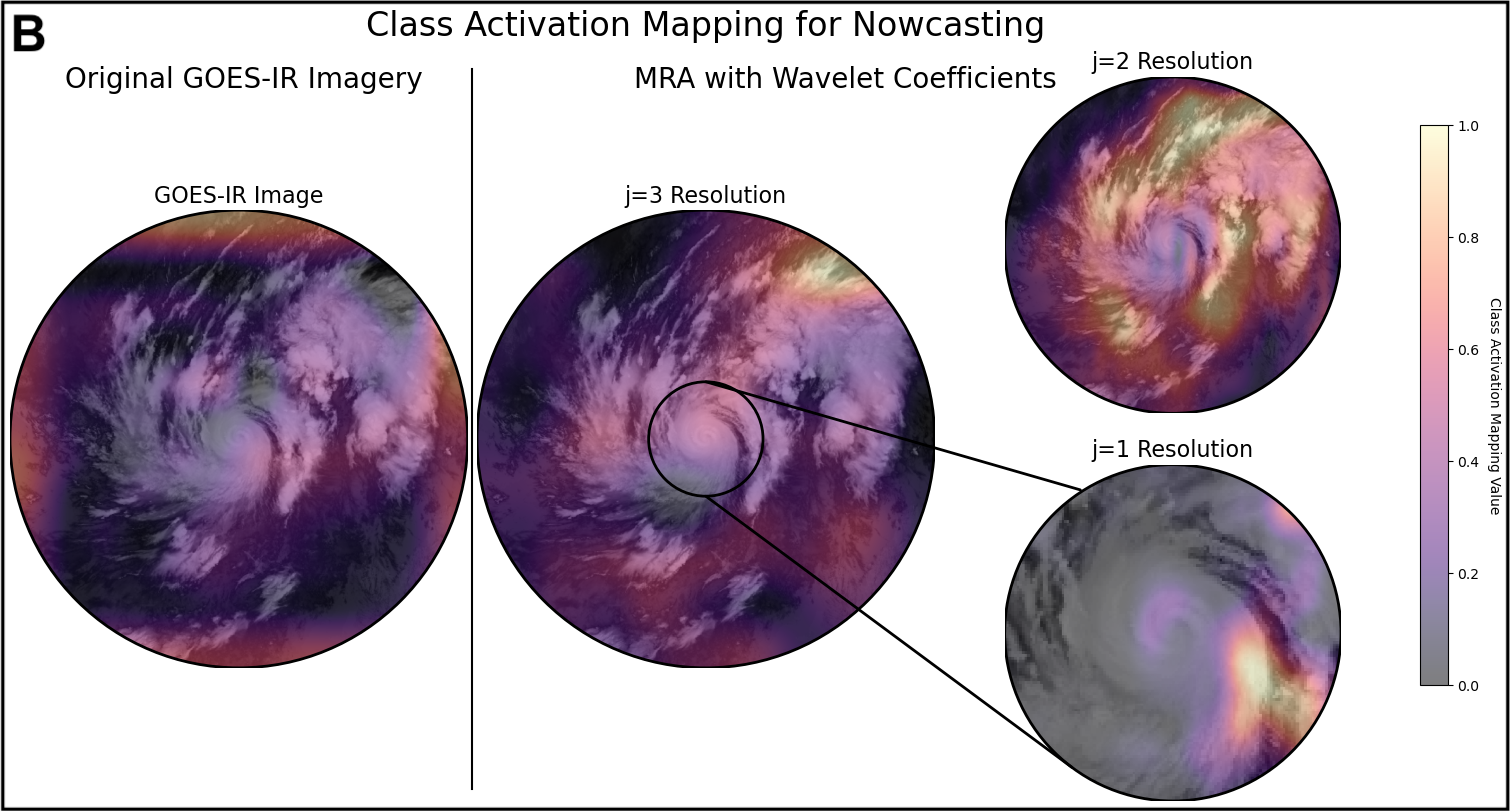}
    \caption{\small {\em \textbf{Panel A:}} Example images at time $t=0$ (i.e., the last frame in a sequence $\mathbf{S_{\leq 0}}$) that is inputted to the respective CNNs to nowcast TC intensities for Hurricane Eta (2020) which underwent RI at $t=0$. {\em Left:} The high-resolution GOES-IR imagery. {\em Right: } The thresholded wavelet coefficients used as model inputs. In this case, $j=3$ represents the coarsest scale, and $j=1$ represents the finest scale used. {\em \textbf{Panel B:} }Example of class activation maps (CAMs) for the inputs at $t=0$ seen in panel A. {\em Left:} The CNN trained on the original GOES-IR imagery had a hard time discerning structural aspects of the storm, instead applying weight at the input's edges. {\em Right:} For the wavelet approach, the CNN was able to identify unique and specific structural aspects of the storm correlated with intensification, such as the storm's curvature and cloud coverage. Specific convective features were identified at all resolutions.}
    \label{fig:t1}
\end{figure}

Next we train a convolutional neural network (CNN) to classify 24-hour sequences $S_{\leq t}$ of wavelet images as structurally primed for an RI event ($Y=1$) or a non-RI event ($Y=0$). For comparison, we also train a similar CNN classifier on the original GOES high-resolution image. We assign each sequence a score $p_t = \mathbb{P}(Y_t=1|S_{\leq t}) - \mathbb{P}(Y_t=1)$, which compares the probability of a sequence $S_{\leq t}$ leading up to an RI event at time $t$ with the marginal probability of such an event. If $p_t > 0$, we label the sequence $S_{\leq t}$ as structurally primed for RI. The early results show a true positive rate of approximately 80.77\% for the MRA approach. When incorporating SHIPS environmental variables into the NN architecture, this true positive rate increases to 84.62\%. The receiver operating characteristic (ROC) curve for the wavelet coefficients and the GOES images with and without the SHIPS variables are shown in Figure \ref{fig:t0}. A benefit of using wavelet coefficients over the high-resolution GOES image is enhanced interpretability, data compression, and robustness to spurious noise. Figure \ref{fig:t1} shows that it is easier to identify key structures associated with a label of RI or non-RI through a class-activation map~\cite{CAM, Captum} in the wavelet space.

\subsection{Deep Learning Architecture for Structural Forecasting}
 Accurate structural forecasts are imperative both to improve 24-hour intensity forecasts, and to allow human forecasters to relate intensity predictions to physical processes. Our full proposed framework is to generate 24-hour structural forecasts in the wavelet space, and then apply the nowcasting framework of TC intensities, outlined in Section 3.2, to support more accurate TC intensity forecasts at a 24-hour lead time. The sparse nature of the wavelet coefficients allows for more efficient NN architectures with generative models. We propose building an encoder-decoder architecture based on transformer blocks with multi-head attention for autoregressive forecasting of wavelet coefficients at future time steps~\cite{Attention}. By discretizing the wavelet coefficients into a discrete vocabulary based on distribution, we can apply positional embeddings to tokenized inputs to effectively learn complex temporal dependencies~\cite{WaveletProp}. Our proposed framework combines the power of deep learning networks to propagate sequences of images and nowcast TC intensities with the invaluable expertise of human forecasters.

\printbibliography[title = {References}]
\end{document}